# Reynolds' Dream?

## A Description of Random Field Theory within a Framework of Algebraic Analysis and Classical Mechanics.

*You can never solve a problem on the level on which it was created.*
*-- Albert Einstein (1879-1955)*

Jerzy Hańćkowiak
Department of Mechanical Engineering, University of Zielona Góra.
ul. Podgórna 50, PL 65-246, Zielona Góra Poland.
E-mail: J.Hanckowiak@ibem.uz.zgora.pl

**ABSTRACT**

Equations for correlation functions, here referred to as Reynolds- Kraichnan- Lewis equations (RKLE), are considered and their wide application is indicated. Perturbation and non-perturbation solutions are given. To elucidate a closure problem - various forms of equations are presented. Exact, closed equations for the projected correlation functions are derived ($\varphi^3, \varphi^4$-models).

## CONTENT OF THE PAPER

In **Sec. 2** Eqs (1.4) are expressed in an equivalent vector form.

In **Sec. 3** general solutions to the RKLE in the free Fock space are constructed.

In **Sec. 4** permutation symmetry is taken into account.

In **Sec. 5** the Aristotelian and Newtonian evolutions are prescribed to RKLE.

In **Sec. 6** constraints upon evolutions are imposed and the d'Alembert principle formulated. In Sec. 6.2 we explain what we mean by the ambiguous phrase - *to solve a closure problem*.

In **Sec. 7** the closed equations for the lowest *n*-pfs are discussed.

Finally, in **Sec. 8** we look at the results from a broader point of view including the renormalization group perspective, the role of information loss, and non- symmetrical n-pfs.

# 1. INTRODUCTION

Despite the growing calculation potential of computational methods in dealing with ever-increasing demands for new technological solutions, basic problems still remain at the level of solutions to non-linear equations. For instance, satisfactory solutions have still not been found for all scales of application of the Navier-Stokes equations in fluid mechanics

$$\rho \frac{d\vec{v}}{dt} = -\nabla p + \eta \Delta \vec{v} + \left(\varsigma + \frac{\eta}{3} grad\, div\vec{v}\right)$$

(0.1)

even in the case where, for a non-compressible fluid $(div\vec{v} = 0)$, one can eliminate the pressure $p$ from (0.1) to get

$$\frac{\partial}{\partial t} rot\,\vec{v} = rot[\vec{v}\, rot\,\vec{v}] + \nu \Delta\, rot\,\vec{v}$$

(0.2)

Landau et al. (1988; page 74). In Gad-el-hak (1998) we read: "it may not be possible to simulate very high Reynolds ($Re = U_0 L \rho / \mu$) number turbulent flows any time soon".

To explore the kernel of difficulty, the problem is simplified to a nonlinear equation of a one scalar field

$$L\varphi(x) = -\lambda \varphi(x)^2 - \mu \varphi(x)^3 \equiv N(\varphi(x))$$

(1.1)

with a linear operator $L$ which is a right invertible operator (e.g., wave operator) and $x = (t, \vec{x}) \in T \times \Omega \subset R \times R^3 = R^4$. For a characterization of all the solutions to equations like (1.1), see Esquivel-Avila (2003). For a rigorous description of the class of right invertible linear operators $L$, see Przeworska-Rolewicz (1988). For a direct application of equations like (1.1) in fluid dynamics, see Burgers equation, Roache (1976).

The general solution to Eq. (1.1) is a function depending on the time-space vector $x$ and a collection of functions $\alpha, \beta$ describing initial and boundary conditions (IC, BC)

$$\varphi = \varphi[x; \alpha, \beta]$$

(1.2)

The mean values of this solution are generally used, however, because they behave more regularly and are more practical than the equations themselves, Frost et al. (1977). Unfortunately, when we introduce averaged quantities, a virtual Pandora's box is opened. In order to close equations for averages, instead of a one-averaged field, we have to consider an infinite number of averaged products of the field (1.2) or to find an appropriate n-point function (n-pf) (moment)

$$V(x_{(n)}) = i^n \int \varphi[x_1; \alpha, \beta] \cdots \varphi[x_n; \alpha, \beta] \cdot P[\alpha, \beta] \delta\alpha \cdot \delta\beta$$

(1.3)

with a given probability density $P$. These are the *functional integrals* - the construction of which involves two formidable tasks : first, the general solution to the nonlinear equation (1.1) has to be obtained, second, the infinitely multiple integral (with only heuristic meaning) has to be executed.

A number of approaches to this difficulty have been proposed. Some have suggested, for example, that entities like (1.3) are non-computable quantities. This need not contradict the opinion that all physical laws are computable, Penrose (1994), but may only mean that the functional integrals are not suitable for all values of constants like $\lambda, \mu$. Other more optimistic approaches include prescribing a number to a given functional integral (1.3) by carrying out perturbation expansion of the general solution (1.2) and expanding the density *P* in the vicinity of the Gaussian density, for instance. For comparison, see Rzewuski (1969), where similar integrals in Quantum Field Theory are considered. A number can also be prescribed to (1.3) using non perturbation methods, see 't Hooft, G. 2002$_a$. In the present paper, a meaning is assigned to (1.3) by means of various equations satisfied by (1.3).

These equations are derived by using (1.1) and assuming that the linear operator *K* commutes with the functional integration of (1.3) and by treating the vectors $x_1,...,x_n$ as independent variables. We thus get the following linear equations for the moments (1.3):

$$L_1 V(x_{(n)}) - i\lambda \cdot V(x_1, x_1, x_2,...,x_n) - \mu \cdot V(x_1, x_1, x_1, x_2,...,x_n) = 0$$

(1.4)

for *n=1,2,…*, where $L_1$ - means here the linear operator *L* acting on the variable $x_1$. We call these equations **the Reynolds-Kraichnan-Lewis equations** (RKLE), Frost et al. (1977), Kraichnan et al. (1962), Monin et al. (1967).

We shall get the same equations if instead of averaging (1.3) we use the convolutions described in Sec.8, where a modification of RKLE is discussed.

To find the n-pfs

$$V(x_{(n)}) \equiv V(x_1,...,x_n)$$

(1.5)

we do not need to know the general solution (1.2), but we have to be able to solve the infinite chain of Eqs (1.4) at IC and BC (IBC) imposed, this time, on the n-pfs $V(x_{(n)})$.

Equations in which IBC as well as the sources (inhomogeneous terms) and coefficients of equations are random often arise due to a reduction of the original 3-D problem by a simplified 1- or 2-D engineering theory in which external surface forces are no longer described by appropriate BC. In other words, starting with a more fundamental formulation of the problem, one can reduce stochastic factors of a theory to the random IBC only. Hence, RKLE, supply a very general description of stochastically excited nonlinear dynamic systems which are often encountered in practice (constructions vibrating in the wind, nonlinear suspensions in vehicles on random road surfaces or structures excited by wave motions at sea).

RKLE have a number of important advantages and applications:
- they can be applied to numerical simulation techniques where random quantity errors appear, Berry at al. (2005).

- they may help to define the functional integrals (1.3), where, in more than four dimensions, difficulties are encountered even with heuristic definitions, 't Hooft (2002$_a$).

- they can even be used in cases where Eq. (1.1) is satisfied by the quantum field $\hat{\varphi}$ fulfilling canonical commutation relations, Weinberg (1995). For that purpose we need to introduce additional time (*quantum time*) and consider stationary correlation functions (1.3) with respect to the quantum time, Hańćkowiak (1992).

- they can be considered as a linearization of the original, nonlinear Eqs (1.1), but unlike the equivalent linearization analysis, Rizzi et al. (2002), this linearization is exact. The cost that has to be paid for this is the resulting infinite system of equations (1.4), for moments (1.3). The term - the *closure problem* – has been coined to express difficulties occurring, Monin et al. (1967). In the present paper, this problem is analysed from the perspective of the free Fock space.

Solutions to Eqs (1.4) expressed by (1.3) should be symmetric with respect to permutations of vectors $x_1,...,x_n$:

$$V(x_1,...,x_n) = V(x_{j1},...,x_{jn})$$

(1.6)

The last property has had long range repercussions in dealing with n-pfs and is intensively used in any process of constructing approximated solutions to Eqs (1.4), see Rzewuski (1969), Vasiliev (1976). In this paper, however, we use a formalism in which conditions (1.6) are not treated as an indispensable limitation of the theory, but as a condition which may or may not be satisfied, whereas from the above perspective - the *symmetry conditions* (1.6) introduced at the initial stage of the theory - are seen rather as an obstacle than an advantage in a description.

## 1.1 GENERATING VECTORS

In the *canonical description* of equations for n-pfs, a crucial role is played by the generating vector

$$|V> \equiv \sum_n \int dx_{(n)} V(x_{(n)}) \hat{\eta}^*(x_1) \cdots \hat{\eta}^*(x_n) |0>$$

(1.7)

for n-pfs $V(x_{(n)})$ with the obvious notation $dx_{(n)} = dx_1 \cdots dx_n$, where an *auxiliary operator valued field* $\hat{\eta}^*(x)$ commutes at different points $x, y$:

$$[\hat{\eta}^*(x), \hat{\eta}^*(y)] = 0$$

(1.8)

In the canonical description, the generating vectors are simply functionals and the vector |0> is represented by the constant functional equal to the number 1, Rzewuski (1969). Acting upon the generating functional (1.7) with a functional derivative of and appropriate order, at $\eta = 0$, one can retrieve all permutationally symmetric *n*-pfs.

In this paper, we use generating vectors (1.7) with the auxiliary operator field $\hat{\eta}^*(x)$ which does not commute at different points $x, y$. Instead, we only postulate the (co) *Cuntz relations*

$$\hat{\eta}(x)\hat{\eta}^*(y) = \delta(x-y) \cdot \hat{I}$$

(1.9)

where $\delta$ is the 4D Dirac function in $R^4$ and $\hat{I}$ is the unit operator in *F*, see below. In fact, in the Cuntz relations, the star denoting Hermitian conjugation is over the left operator, Cuntz (1977).

Using the quantum field terminology, the auxiliary operator field $\hat{\eta}(x)$ "annihilates" the "vacuum" vector |0> :

$$\hat{\eta}(x)|0> = 0$$

(1.10)

The linear space of vectors (1.7) with relations (1.9-10) is called *the free (Boltzmanian) Fock space* and is denoted by $F$, see Greenberg (2000), Accardi (1999). In the case of the free Fock space relationships among products $\hat{\eta}(x_1)\cdots\hat{\eta}(x_n)$ are not imposed, hence the term "free", Fried (1979). To some extent, the vectors (1.7) with (1.8-9) remind us of the free vectors of affine geometry, Spindler (1994 vol.1).

The linear field operators $\hat{\eta}, \hat{\eta}^*, \hat{I}$, and other operators constructed with their help, act in the space $F$, Sec.2.3.

In the case of discrete variables *x,y*, the Dirac delta in (1.9) is substituted by the Kronecker delta. In this case, at every space-time point *x,* the auxiliary field $\eta(x)$, is a *co-isometry operator*. We can say that the auxiliary field introduced is an *co-isometry operator field*. At this point we would like to stress that the operator valued auxiliary fields $\hat{\eta}$ are used to obtain a formalism which is related to other branches of mathematics like Algebraic Analysis or Idempotent Analysis, in which, at least for certain class of problems, general solutions can be relatively easily found, see Przeworska-Rolewicz (1988) and Kolokoltsov et al. (1997).

## 1.2 DIRAC'S NOTATION

We use Dirac's notation (1964) in which bra and ket vectors are considered. Vectors bra, <B|, can be represented by one row matrices and ket vectors, |A>, by one column matrices (all infinitely long). They form a pair of dual spaces denoted by $F^d$ and $F$

In this notation the scalar product of vectors $|A>, |B> \in F$ is denoted by $<A|B>$ and has the property

$$<B|A> = \overline{<A,B>}$$

(D.1)

where $\overline{\phantom{x}}$ denotes complex conjugation.

The *ket vector conjugated to the bra vector* $<P|\hat{\alpha}$ is the vector $\hat{\alpha}^*|P>$ in which the conjugate operator $\hat{\alpha}^*$ to the operator $\hat{\alpha}$ appears. We have also

$$<B|\hat{\alpha}^*|P> = \overline{<P|\alpha|B>}$$

(D.2)

## 2. FREE VECTOR FORM OF RKLE

### 2.1 BASIS VECTORS
With the idea of a generating vector, we can regard Eqs (1.4), for n-pfs $V(x_{(n)})$, as equations for *components* of the vector |V>. With the basis vectors $\hat{\eta}^*(y_1)\cdots\hat{\eta}^*(y_m)|0>$, we obtain from (1.7) and (1.9-10)

$$<0|\hat{\eta}(y_1)\cdots\hat{\eta}(y_m)|V> = V(y_{(n)})$$
(2.1)

where we took into account the property

$$<0|\hat{\eta}^*(x) = 0,$$
(2.2)

resulting from (1.10), and that

$$<0|\hat{\eta}(y_1)\cdots\hat{\eta}(y_n)\hat{\eta}^*(x_1)\cdots\hat{\eta}^*(x_n)|0> = \delta(y_1 - x_1)\cdots\delta(y_n - x_n)$$
(2.3)

In the case of the commuting auxiliary field $\hat{\eta}^*(x)$, the action of $\hat{\eta}(x)$ would be substituted by the action of the functional derivative at zero function, Rzewuski (1969). Then, on the r.h.s. of (2.3), the *n*! terms resulting from permutationally symmetric basis would appear, e.g., for *n=11, n!=39 916 800* terms!, Hańćkowiak (1993).

### 2.2 ELEMENTARY OPERATORS AND THE VECTOR FORM OF RKLE (1.4)
With the help of operators $\hat{\eta}$ one can express the unit operator acting in the space *F* of vectors (1.7) with (1.9-10) as follows:

$$\hat{I} = \int\hat{\eta}^*(x)\hat{\eta}(x)dx$$
(2.4)

In fact, this is the unit operator in the space not containing the vacuum vector |0>. In the complete space, the unit operator (2.4) should be substituted by the operator

$$\hat{I} = \int\hat{\eta}^*(x)\hat{\eta}(x)dx + |0><0|$$

Introducing operators expressed by the formal integrals

$$\hat{L} = \int\hat{\eta}^*(x)L(x,y)\hat{\eta}(y)dxdy$$
(2.5)

$$\hat{N} = -i\lambda\int\hat{\eta}^*(x)\hat{\eta}(x)^2 dx - \mu\int\hat{\eta}^*(x)\hat{\eta}(x)^3 dx \equiv -i\lambda\hat{N}_1 - \mu\hat{N}_2$$
(2.6)

one can describe Eqs (1.4) in an equivalent, operator form

$$\left(\hat{L} + \hat{N}\right)|V> = 0$$
(2.7)

Indeed, it is easy to see that from (2.7), with the help of non-commuting operators (1.9) and the non-symmetrical basis vectors satisfying (2.3), we obtain equations (1.4). In other words, to derive RKLE we do not need permutation symmetry. This means that RKLE can be considered in a larger space *F* incorporating permutationally symmetrical and non-symmetrical n-pfs.

## 2.3 IS THERE AN ANALOGUE TO THE FUNCTIONAL DERIVATIVE IN *F*?

Below, we show that $\hat{L}$ is a linear diagonal operator and $\hat{N}_{1,2}$ are linear upper triangular operators. These last two operators are responsible for the closure problem. Using the anti-Cuntz relations (1.9) one can show that

$$\hat{N}_1^2 = \hat{N}_2$$

and a more general equation

$$\hat{N}_1^n = \hat{N}_n$$

(2.8)

where the operators

$$\hat{N}_n := \int \hat{\eta}^*(x)\hat{\eta}(x)^{n+1} dx; \quad n = 1,2,3...$$

(2.9)

These operators can be used for the description of a more general polynomial nonlinear theories than (1.1). It is interesting that if in (2.9) we put *n=0*, then we get the unit operator (2.4). The case *n=1* corresponds to the second functional derivative and so on. In other words, in the free Fock space there is no analogue to the first order of the functional derivative.

## 2.4 PROJECTION PROPERTIES OF OPERATORS

Due to the presence of the operator $\hat{N}$, equations (2.7) or (3.8) generate an infinite chain of equations for n-pfs $V(x_{(n)})$. This operator is an *upper triangular operator* with respect to the Hermitian projectors $\hat{P}_n$, which, by definition, from the vector $|V>$ exclude those parts containing the appropriate n-pfs:

$$\hat{P}_n |V> = \int V(x_{(n)})\hat{\eta}^*(x_1)\cdots\hat{\eta}^*(x_{(n)})|0> dx_{(n)}$$

(2.10)

The projectors $\hat{P}_n$ can be expressed by means of the operators $\hat{\eta},\hat{\eta}^*$ satisfying relations (1.9) and (1.10) using the tensor (diadic) products of the basis vectors:

$$\hat{P}_n = \int \hat{\eta}^*(y_1)\cdots\hat{\eta}^*(y_n)|0><0|\hat{\eta}(y_n)\cdots\hat{\eta}(y_1) dy_{(n)}$$

(2.11)

These are Hermitian projectors satisfying the orthonormality relations

$$\hat{P}_n\hat{P}_m = \delta_{nm}\hat{P}_n$$

(2.12)

and the completeness equality

$$\sum_{n=1}^{} \hat{P}_n = \hat{I}$$

(2.13)

see (2.4). It turns out that $\hat{N}$ is an *upper triangular* operator

$$\hat{P}_n\hat{N} = \hat{P}_n\hat{N}(\hat{P}_{n+1} + \hat{P}_{n+2})$$

(2.14)

The operator $\hat{L}$ is *diagonal*

$$\hat{P}_n\hat{L} = \hat{L}\hat{P}_n$$

(2.15)

for *n=1,2*.... We also have

$$\hat{P}_1\hat{N} = \hat{0}, \quad \hat{P}_0\hat{L} = \hat{0}$$

(2.16)

where projector

$$\hat{P}_0 = |0><0|$$

(2.17)

## The main assumptions

In fact, a linear operator is a pair $(A, dom\, A)$, because it is determined by its mapping and its domain of definition. We assume that the domains of all the operators used are the same. By making this assumption we hope that, where rigorous mathematics is not the main aim of our considerations, we are not flouting the standards of scientific research. Our ultimate purpose is to derive, with the help of feasible assumptions, equations which can be effectively solved and whose solutions can be confronted with the original equations and with experimental results.

## 3. GENERAL SOLUTION IN FREE FOCK SPACE

### 3.1 RIGHT INVERTIBLE OPERATORS. DISCRETE SPACE-TIME.

If the *kernel function L(x,y)* of the operator $\hat{L}$ is a *right invertible*

$$\int L(x,y) L_R^{-1}(y,z) dy = \delta(x-z)$$

(3.1)

then the operator $\hat{L}$ is a right invertible operator

$$\hat{L}\hat{L}_R^{-1} = \hat{I}$$

(3.2)

with a right inverse operator

$$\hat{L}_R^{-1} = \int \hat{\eta}*(\tilde{y}) L_R^{-1}(\tilde{y},z) \hat{\eta}(z) d\tilde{y} dz$$

(3.3)

We denote right and left inverses by additional sub indexes *R,L*.

**Right invertible operators** are very common in science and they express an important property of linear systems, namely that various physical situations which usually correspond to source-free fields may correspond to a given source of the field. The importance of this concept was expressed in the Przeworska-Rolewicz's book: - Algebraic Analysis - which is "the theory of right invertible operators in linear space (without any topology, in general)", (1988; page XIV). It is astonishing that in the Reynolds-Kraichnan-Lewis approach to Eqs (1.1) even operators related to their non-linear parts are right invertible.

We look for right inverses to the operators $\hat{N}_{1,2}$ in a form

$$\left(\hat{N}_1\right)_R^{-1} = \int N_{1R}^{-1}(y_{(3)})\hat{\eta}*(y_1)\hat{\eta}*(y_2)\hat{\eta}(y_3) \cdot dy_{(3)}$$

(3.4)

and

$$\left(\hat{N}_2\right)_R^{-1} = \int N_{2R}^{-1}(y_{(4)})\hat{\eta}*(y_1)\hat{\eta}*(y_2)\hat{\eta}*(y_3)\hat{\eta}(y_4) \cdot dy_{(4)}$$

(3.5)

From the definition

$$\hat{N}_{1,2}\left(\hat{N}_{1,2}\right)_R^{-1} = \hat{I}$$

(3.6)

and from, for example, (3.5)

$$N_{2R}^{-1}(x,x,x,y) = \delta(x-y)$$

(3.7)

a restriction satisfied by many 4-pfs. For example, a 4-pf

$$N_{2R}^{-1}(y_{(4)}) = f(y_1 - y_2)\delta(y_3 - y_4)$$

with $f(0)=1$. In the discrete case, we can choose

$$N_{2R}^{-1}(y_{(4)}) = \delta_{y1y2}\delta_{y3y4}$$

with the 4D Kronecker delta instead of the Dirac delta function. In this case,

$$(\hat{N}_{1,2})_R^{-1} = \hat{N}_{1,2}*$$

(3.8)

and we see that the operators $\hat{N}_{1,2}$ responsible for the closure problem in RKLE (1.4), are not only right invertible but are also the *co-isometry* operators.

**3.2 THE PERTURBATION SERIES SOLUTION AND INITIAL OPERATORS**

Multiplying (2.7) by the operator $\hat{L}_R^{-1}$ we get

$$\left(\hat{I} + \hat{L}_R^{-1}\hat{N}\right)|V> = \hat{P}_L |V>$$

(3.9)

where the operators

$$\hat{P}_L \equiv \hat{I} - \hat{L}_R^{-1}\hat{L} \equiv \hat{I} - \hat{Q}_L$$

(3.10)

are *projectors* (idempotent operators)

$$\hat{P}_L = \hat{P}_L^2, \quad \hat{Q}_L = \hat{Q}_L^2, \quad \hat{P}_L + \hat{Q}_L = \hat{I}$$

(3.11)

$$\hat{L}\hat{P}_L = \hat{0}, \quad \hat{L}\hat{Q}_L = \hat{L}$$

All relations (3.11) result from the definition of the projector $\hat{Q}_L \equiv \hat{L}_R^{-1}\hat{L}$. One can show that $\hat{P}_L$ is a projector upon the whole null space of the operator $\hat{L}$ and it is seen directly from (3.10) that (3.11) takes place. The projector $\hat{P}_L$ is called an *initial operator for an operator L corresponding to a right inverse* $\hat{L}_R^{-1}$ *of* $\hat{L}$, Przeworska-Rolewicz (1988). The projector $\hat{Q}_L$ is its completion to the unit operator. We can also say that the projector $\hat{P}_L$ projects the free Fock space $F$ onto the null space of the operator $\hat{L}$ in the *direction* $\hat{Q}_L F$. The direction depends on the choice of right inverse operator $\hat{L}_R^{-1}$.

From Eqs (3.9) and (3.10) we get

$$\hat{L}_R^{-1}(\hat{L} + \hat{N})|V> = 0$$

which acted on by the operator $\hat{L}$ gives the original Eq. (2.7), see (3.2). This means that **Eqs (2.7) and (3.9-10) are equivalent to each other**.

Eq. (3.9) can be used to represent a formal solution to Eq. (2.7) in the power series (Neumann series)

$$|V> = (\hat{I} + \hat{L}_R^{-1}\hat{N})^{-1} = \sum_{n=0}(-1)^n (\hat{L}_R^{-1}\hat{N})^n \hat{P}_L |V>$$

(3.12)

Moreover

$$\hat{P}_L \hat{L}_R^{-1} = 0$$

(3.13)

So, if we act on Eq. (3.12) with projector $\hat{P}_L$ then we get identity. This means that (3.12) with an **arbitrary projection** $\hat{P}_L |V>$ represents the general solution to Eq. (2.7). We have to remember, however, that (3.12) is considered in the free Fock space $F$ and hence, for arbitrary choice of $\hat{P}_L |V>$, (3.12) may generate symmetrical and non-symmetrical solutions.

## 3.3 BEYOND THE PERTURBATION SERIES

Admitting another assumption, namely that operator $(\hat{I} + \hat{L}_R^{-1}\hat{N})$ in Eq. (3.9) is also a right invertible:

$$(\hat{I} + \hat{L}_R^{-1}\hat{N})(\hat{I} + \hat{L}_R^{-1}\hat{N})_R^{-1} = \hat{I}$$

(3.14)

and introducing projectors

$$\hat{Q} = (\hat{I} + \hat{L}_R^{-1}\hat{N})_R^{-1}(\hat{I} + \hat{L}_R^{-1}\hat{N}) \equiv \hat{I} - \hat{P}$$

(3.15)

we can express a solution to Eq. (3.9) as

$$|V> = (\hat{I} + \hat{L}_R^{-1}\hat{N})_R^{-1}\hat{P}_L |V> + \hat{P}|V>$$

(3.16)

where on the r.h.s. of (3.16), the vector $\hat{P}_L |V>$ belonging to the null space of the operator $\hat{L}$ and the vector $\hat{P}|V>$ belonging to the null space of the operator $\left(\hat{I}+\hat{L}_R^{-1}\hat{N}\right)$ appear. In fact, from the point of view of the free Fock space, these two projections of the vector $|V>$ can again be **arbitrarily chosen**, Przeworska-Rolewicz (1988). However, from (3.13) and (3.14)

$$\hat{P}_L\left(\hat{I}+\hat{L}_R^{-1}\hat{N}\right)_R^{-1} = \hat{P}_L$$

(3.17)

and from (3.16) results

$$\hat{P}_L \hat{P}|V> = 0$$

(3.18)

This is of course satisfied for an unique solution to Eq. (3.9), because then the operator $\hat{P}=\hat{0}$. If a right inverse in the formula (3.16) can be approximated by the unit operator, we have

$$|V> \approx \hat{P}_L|V> + \hat{P}|V>$$

(3.19)

In this case (3.18) tells us that the component $\hat{P}_L|V>$ related to the linear part of the theory belongs to a different subspace from the component $\hat{P}|V>$ related to the non-linear part. By definition, Eqs (2.7) with property (3.14) are classified as *ill-determined,* Pogorzelec (1983), Przeworska-Rolewicz (1988).

### 3.4 NON-PERTURBATIVE, ALMOST EXPLICIT SOLUTIONS

In the free Fock space (1.7) with (1.9-10), there is a huge amount of (almost) explicit solutions to Eq. (2.7) including, according to formulas (3.18-9) those which are drastically different from solutions with small values of the constants $\lambda, \mu$.

To construct this set of solutions let us multiply Eq. (2.7) by a right inverse $\hat{N}_R^{-1}$ to the operator $\hat{N}$. We get an equivalent equation to Eq. (2.7)

$$\left(\hat{N}_R^{-1}\hat{L}+\hat{I}\right)|V> = \hat{P}_N|V>$$

(3.20)

with a projector

$$\hat{P}_N := \hat{I} - \hat{N}_R^{-1}\hat{N}$$

(3.21)

projecting $F$ onto the null space of the operator $\hat{N}$ in the direction $\hat{Q}_N = \hat{N}_R^{-1}\hat{N}_R$, Przeworska-Rolewicz (1988). The general solution to (3.20) or (2.7) is

$$|V> = \left(\hat{I}+\hat{N}_R^{-1}\hat{L}\right)^{-1}\hat{P}_N|V> \Leftrightarrow \hat{P}_n|V> = \sum_{k=0}^{finite}(-1)^k \hat{P}_n(\hat{N}_R^{-1}\hat{L})^k \hat{P}_N|V>$$

(3.22)

with the inverse to the diagonal plus the lower triangular operator, Sec.2, and the **arbitrary projection vector** $\hat{P}_N|V>$. The projectors $\hat{P}_n$ are defined by (2.11).

From definition (3.21)

$$\hat{P}_N \hat{N}_R^{-1} = \hat{0}$$

(3.23)

So, $\hat{P}_N$ is an initial operator for the operator $\hat{N}$ corresponding to a right inverse $\hat{N}_R^{-1}$ of $\hat{N}$. $\hat{P}_n |V>$ in (3.22) can be represented by a finite power series, because the product $\hat{N}_R^{-1} \hat{L}$ is a lower triangular operator.

The problem which we have to face in the next sections is the following: the solutions represented by formula (3.12) are given by the projections $\hat{P}_L |V>$ of the vector $|V>$ which are related to physical quantities (n-pfs and their first derivatives at the initial time), but the series (3.12) contains, for every n-pf $V(x_{(n)})$, infinite terms. The solutions given by formula (3.22), for every n-pf $V(x_{(n)})$ contain a finite number of terms, but the n-pfs generated by the projections $\hat{P}_N |V>$ are not directly related to known quantities.

## 4. PERMUTATION SYMMETRY

Up to now we have ignored the permutation symmetry of n-pfs which in fact leads to many restrictions upon them. In this section we show how it is possible in the free Fock space to take into account this important property of n-pfs.

### 4.1 PERTURBATION THEORY

To get a permutationally symmetric solution from the general solution created in Sec.3 - we must, in some way, restrict the arbitrary projection vectors entering formulas (3.12) or (3.22). This restriction can be achieved by introducing a projector $\hat{S}$ upon the permutation symmetric n-pfs. In other words, we look for *symmetrical* generating vectors satisfying the condition

$$|V> = \hat{S}|V> \Leftrightarrow (\hat{I} - \hat{S})|V> = 0$$

(4.1)

where the Hermitian projector

$$\hat{S} = \sum_{n,perm} \frac{1}{n!} \int \hat{\eta}^*(y_{i1})...\hat{\eta}^*(y_{in})|0><0|\eta(y_n)...\eta(y_1) dy_{(n)}$$

(4.2)

Let us present the formal, general solution (3.12) in a form

$$|V> = (\hat{I} + \hat{L}_R^{-1}\hat{N})^{-1} \hat{P}_L |V> = (\hat{I} + \hat{L}_R^{-1}\hat{N})^{-1} (\hat{P}_L |V>^{(0)} + \hat{P}_L |V>^{(1)})$$

(4.3)

where for $\lambda = \mu = 0$

$$|V> = |V>^{(0)} = \hat{P}_L |V>^{(0)} = \hat{S}|V>^{(0)}$$

(4.4)

and for $\lambda \neq 0, \mu \neq 0$

$$|V> = |V>^{(0)} + |V>^{(1)}$$

(4.5)

For an arbitrary choice of projections $\hat{P}_L |V>^{(0,1)}$, (4.3) represents a superposition of the two solutions on Eq. (2.7). We would like to choose these arbitrary projections of the formula (4.3) such that condition (4.1) is satisfied. We get

$$\hat{Q}_L^S \hat{P}_L |V> = \hat{Q}_L^S \left( \hat{P}_L |V>^{(0)} + \hat{P}_L |V>^{(1)} \right) = 0 \tag{4.6}$$

where the projector (idempotent operator)

$$\hat{Q}_L^S := \left( \hat{I} + \hat{L}_R^{-1} \hat{N} \right) \left( \hat{I} - \hat{S} \right) \left( \hat{I} + \hat{L}_R^{-1} \hat{N} \right)^{-1} \tag{4.7}$$

Introducing the complementary projector

$$\hat{P}_L^S := \left( \hat{I} + \hat{L}_R^{-1} \hat{N} \right) \hat{S} \left( \hat{I} + \hat{L}_R^{-1} \hat{N} \right)^{-1}; \quad \hat{P}_L^S + \hat{Q}_L^S = \hat{I} \tag{4.8}$$

we can say that in symmetrical theories, the arbitrary vector $\hat{P}_L |V>$ decomposed into two components by means of the complementary projections $\hat{P}_L + \hat{Q}_L = \hat{I}$ has only the component $\hat{P}_L^S \hat{P}_L |V> \neq 0$. Taking into account (4.5), we additionally postulate

$$\hat{P}_L^S \hat{P}_L |V>^{(1)} = 0 \tag{4.9}$$

This assumption can be justify as follows. We must bear in mind that in the free Fock space the projection $\hat{P}_L |V>^{(1)}$ is an arbitrary vector from the null space of the operator $\hat{L}$. The symmetry condition (4.6) has allowed us to restrict only one component of the projection $\hat{P}_L |V>^{(1)}$. Component (4.9) l was still arbitrary and we made it equal to zero. It is well known that even in symmetrical theory, subsequent approximations of perturbation theory are not uniquely determined, see e.g. Bobolubov (1976; page 169). So, (4.9) can be interpreted as a restriction imposed on the theory to eliminate the above arbitrariness.

The above assumptions lead to

$$\hat{P}_L |V> = \hat{P}_L^S \hat{P}_L |V>^{(0)} \tag{4.10}$$

Let us insert the unit (4.8) between the inverse and projected vector in formula (4.3). We get

$$|V> = \hat{S} \left( \hat{I} + \hat{L}_R^{-1} \hat{N} \right)^{-1} \left( \hat{P}_L |V>^{(0)} \right) \tag{4.11}$$

In other words, assumptions (4.6) and (4.9) lead to the permutationally symmetric n-pfs. In this theory, the projection vector $\hat{P}_L |V>$, arbitrary in the whole space $F$, is given by (4.10), (4.8) and (4.4). Most surprising is the fact that we get identical results if we use (4.4) and symmetrize the results obtained, see (4.11).

### 4.2 PERMUTATION SYMMETRY OF NON-PERTURBATIVE SOLUTIONS

Now we apply similar consideration to the general solution (3.22) using instead of (4.4) the assumption

$$|V>^{(0)} = \hat{P}_N |V>^{(0)} = \hat{S} |V>^{(0)} \tag{4.12}$$

for $\lambda = \mu = \infty$. We postulate, for $\lambda = \mu \neq \infty$, (4.5), and instead of (4.6) we have

$$\hat{Q}_N^S \left( \hat{P}_N |V>^{(0)} + \hat{P}_N |V>^{(1)} \right) = 0$$

(4.13)

with the projector

$$\hat{Q}_N^S := \left( \hat{I} + \hat{N}_R^{-1} \hat{L} \right)\left( \hat{I} - \hat{S} \right)\left( \hat{I} + \hat{N}_R^{-1} \hat{L} \right)^{-1}$$

(4.14)

(4.12) means that

$$\hat{N} |V>^{(0)} = 0$$

(4.15)

In this case we get

$$|V> = \hat{S}\left( \hat{I} + \hat{N}_R^{-1} \hat{L} \right)^{-1} \hat{P}_N |V>^{(0)}$$

(4.16)

It turns out that (4.15-16) lead to trivial solutions of products of Dirac's deltas and their derivatives. To avoid this result, we propose weakening the above assumptions by using an adapted version of the assumption (4.10):

$$\hat{P}_N |V> = \hat{P}_N^S \hat{P}_N |V>^{(pert)}$$

(4.17)

where the vector $|V>^{(pert)}$ is given by e.g. the perturbation formula recovered from (4.11). We assume here that the perturbation expansion is justified only for the projection (4.17) where the projector

$$\hat{P}_N^S := \left( \hat{I} + \hat{N}_R^{-1} \hat{L} \right)\hat{S}\left( \hat{I} + \hat{N}_R^{-1} \hat{L} \right)^{-1}$$

(4.18)

From (4.17) and (3.22) we get

$$|V> = \hat{S}\left( \hat{I} + \hat{N}_R^{-1} \hat{L} \right)^{-1} \hat{P}_N |V>^{(pert)}$$

(4.19)

How can a formula like (4.19) be justified? If the projection $\hat{P}_N |V>^{(pert)}$ corresponds to a symmetrical theory, we can omit the projector $\hat{S}$ in front of (4.19). Then, it is a trivial task to show that RKLE (2.7) is satisfied.

### 4.3. THE FREE FOCK SPACE $F$ WITH AN INVOLUTION

One can introduce an operator

$$\hat{T}_S = 2\hat{S} - \hat{I}$$

(4.20)

which is an involution operator in the vector space $F$:

$$\hat{T}_S^2 = \hat{I}$$

(4.21)

Spindler (1994). The physical states are eigenvectors of the above involution operator with +1 eigenvalue

$$\hat{T}_S |phys> = |phys> = \hat{S} |phys>$$

(4.22)

A vector from *F* can be presented as

$$|V> = \hat{S}|V> + (\hat{I} - \hat{S})|V> \quad (4.23)$$

and from definition (4.20) of the involution operator $\hat{T}_S$

$$\hat{T}_S |V> = \hat{S}|V> - (\hat{I} - \hat{S})|V> \quad (4.24)$$

We can say that the involution operator $\hat{T}_S$ constructed with the help of the permutation symmetry projector $\hat{S}$ is a *reflection* at $\hat{S}F$ along $(\hat{I} - \hat{S})F$ in the free Fock space *F* decomposed into the direct sum

$$F = \hat{S}F \oplus (\hat{I} - \hat{S})F \quad (4.25)$$

Spindler (1994). The above reflection property takes place for any involution operators related to any projectors used in their definitions, see (4.20). The difference between a projection and reflection can be illustrated as follows:

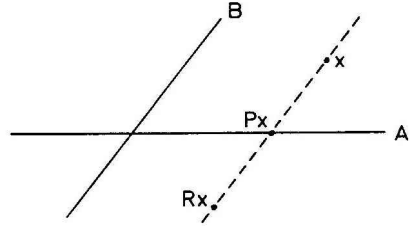

where $A = \hat{S}F$, $B = (\hat{I} - \hat{S})F$, $Px = \hat{S}|V>$, $Rx = \hat{T}_S |V>$.

## 5. ARISTOTELIAN ANE NEWTONIAN EVOLUTIONS

Besides the *"static" equations* (SE) (2.7), (3.9) or (3.20) one can consider various *"evolution" equations* (EE) whose solutions tend asymptotically to solutions of static equations. In this way one can exploit an explicit construction of the general solutions in the free Fock space.

We can use the *Aristotelian* EE,

$$|\dot{V}(s)> = -\beta(s)\left(\hat{A}|V(s)> - |Z>\right) \quad (5.1)$$

with an arbitrary positive function $\beta$ and an appropriately chosen, s-independent, *positive operator* $\hat{A}$

$$<\Psi|\hat{A}|\Psi> \geq 0; \forall |\Psi> \in F$$

and vector $|Z>$. We can use the *Newtonian* EE

$$|\ddot{V}(s)> + \hat{D}(s)|\dot{V}(s)> + \hat{A}|V(s)> - |Z> = 0$$

(5.2)

with a *damping operator D(s)* given by a *positively defined* operator

$$<\Psi|\hat{D}(s)|\Psi>> 0; \forall |\Psi>\in F$$

The dots mean here the time derivatives with respect to the *fictitious time s*. In both cases of EEs, solutions, with $s \to \infty$, tend to solutions of the stationary equation

$$\hat{A}|V> = |Z>$$

(5.3)

This can be justified in a more or less rigorous way, see Vasiliev (1980), where Eqs (5.1-2) are considered for finite dimensional cases such as the *gradient and heavy ball methods*. In the case of the infinite dimensional case, for heuristic justification of the previous statement, see Hańćkowiak (1992) and considerations beginning from formula (5.8).

In the case of Eq. (3.9), the positive operator *A* and the vector |Z> are given by formulas

$$\hat{A} = \left(\hat{I} + \hat{N}*(\hat{L}*)_L^{-1}\right)\left(\hat{I} + \hat{L}_R^{-1}\hat{N}\right); \quad |Z> = \left(\hat{I} + \hat{N}*(\hat{L}*)_L^{-1}\right)\hat{P}_L|V>$$

(5.4)

where the operator

$$(\hat{L}*)_L^{-1} = (\hat{L}_R^{-1})*$$

is a left inverse to the operator $\hat{L}*$. The projection $\hat{P}_L|V>$ entering the vector |Z> is a projection of the vector |V> satisfying the original RKLE (2.7). It is important to notice that Eq. (5.3) with (5.4) is equivalent to Eq. (3.9) or (2.7) because the operator $\left(\hat{I} + \hat{N}*(\hat{L}*)_L^{-1}\right)$ as a sum of the unit and lower triangular operators is non-singular (both side invertible).

## 5.1 HOMOGENOUS FORM OF EVOLUTIONS

Let us describe (5.2) in a homogenous form by introducing a new vector

$$|V(s)> := |\tilde{V}(s)> + \left(\hat{I} + \hat{L}_R^{-1}\hat{N}\right)_G^{-1}\hat{P}_L|V>$$

(5.5)

where the generalized inverse satisfies equation

$$\left(\hat{I} + \hat{L}_R^{-1}\hat{N}\right)\left(\hat{I} + \hat{L}_R^{-1}\hat{N}\right)_G^{-1} = \hat{P}_L$$

(5.6)

This is in fact a projected equation for a right inverse

$$\left(\hat{I} + \hat{L}_R^{-1}\hat{N}\right)\left(\hat{I} + \hat{L}_R^{-1}\hat{N}\right)_R = \hat{I}$$

It is worth noticing that the vector

$$|\Psi> \equiv \left(\hat{I} + \hat{L}_R^{-1}\hat{N}\right)_G^{-1}\hat{P}_L|V>$$

satisfies the original RKLE described in the form (2.7). Indeed, from (5.6) we have $\left(\hat{I} + \hat{L}_R^{-1}\hat{N}\right)|\Psi> \equiv \hat{P}_L |V>$. Multiplying the last equation by $\hat{L}$ and taking into account that $\hat{L}\hat{P}_L = 0$ we get (2.7).

With (5.5), instead of (5.2) we get the *homogenous evolution equation* (HEE)

$$|\ddot{\tilde{V}}(s)> + \hat{D}(s)|\dot{\tilde{V}}(s)> + \hat{A}|\tilde{V}(s)> = 0$$

(5.7)

whose arbitrary solutions, at least for finite dimensional mechanics, tend, for $s \to \infty$, to the trivial solution, Vasiliev (1980). For infinite dimension, see below. In both cases, the vector $|V(s)>$ tends to the *equilibrium vector* $\left(\hat{I} + \hat{L}_R^{-1}\hat{N}\right)_G^{-1}\hat{P}_L|V>$ satisfying Eq. (3.9). We can say that the trivial solution to Eq. (5.7) is the *attractor* with the *basin* constituting the whole space, Nusse et al. (1997).

## 5.2 AN EIGENVECTORS EXPANSION

The attracting properties of evolutions (5.7) can be justified in a more mathematical terms. For that purpose, let us decompose the vector

$$|\tilde{V}(s)> = \sum_k c_k(s)|\Psi_k>$$

(5.8)

with the help of the eigenvectors of the positive operator $A$

$$A|\Psi_k> = a_k|\Psi_k>; \quad a_k > 0$$

(5.9)

From (5.7-9) we get

$$\sum_k \left\{\ddot{c}_k(s) + \dot{c}_k(s)\hat{D}(s) + a_k c_k(s)\right\}|\Psi_k> = 0$$

(5.10)

For a diagonal, constant damping operator

$$\hat{D} = \gamma \cdot \hat{I}; \quad \gamma > 0$$

(5.11)

and from linear independency of the eigenvectors $|\Psi_k>$, we get equations

$$\ddot{c}_k + \gamma \cdot \dot{c}_k + a_k c_k = 0$$

(5.12)

with the general solutions

$$c_k(s) = A_k \exp(\omega_+ s) + B_k \exp(\omega_- s)$$

(5.13)

where

$$\omega_\pm = \frac{-\gamma \pm \sqrt{\gamma^2 - 4a_k}}{2}$$

Hence we see that for $\gamma, a_k > 0$ (the last equality results from the assumption that operator $A$ is positively defined), (5.13) tend to zero, for $s \to \infty$.

## 5.3 AN ENERGY APPROACH

**If appropriate scalar products exist**, we can give an energy justification of the limes above: multiplying the HEE (5.7) by the bra velocity, we get

$$<\dot{\widetilde{V}}|\ddot{\widetilde{V}}> + <\dot{\widetilde{V}}|\hat{D}(s)|\dot{\widetilde{V}}> + <\dot{\widetilde{V}}|\hat{A}|\widetilde{V}> = 0 \tag{5.14}$$

By a similar procedure with conjugate equation (5.7), we get

$$<\ddot{\widetilde{V}}|\dot{\widetilde{V}}> + <\dot{\widetilde{V}}|\hat{D}(s)|\dot{\widetilde{V}}> + <\widetilde{V}|\hat{A}|\dot{\widetilde{V}}> = 0 \tag{5.15}$$

which lead together to

$$\frac{d}{ds}\left(<\dot{\widetilde{V}}|\dot{\widetilde{V}}> + <\widetilde{V}|\hat{A}|\widetilde{V}>\right) = -2<\dot{\widetilde{V}}|\hat{D}(s)|\dot{\widetilde{V}}> \tag{5.16}$$

For the positively defined damping matrix $\hat{D}(s)$, as in the case of $\hat{D} \propto \hat{I}$, the r.h.s. of (5.16) is negative, which means that the expression in the left parenthesis is a decreasing quantity. Using the language of mechanics, we say that the total "energy" of the homogenous system

$$\widetilde{E}(s) \equiv \frac{1}{2}\left(<\dot{\widetilde{V}}(s)|\dot{\widetilde{V}}(s)> + <\widetilde{V}(s)|\hat{A}|\widetilde{V}(s)>\right) \tag{5.17}$$

given by the sum of the *kinetic energy* proportional to the square of the velocity and the *system potential energy* given by the positive operator $\hat{A}$, see (5.4), is decreasing. By integrating (5.16), we get

$$\widetilde{E}(s) - \widetilde{E}(0) = -\int_0^s <\dot{\widetilde{V}}(\vartheta)|\hat{D}(\vartheta)|\dot{\widetilde{V}}(\vartheta)> d\vartheta \equiv \Delta(s) \tag{5.18}$$

Because $\widetilde{E}(s)$ is a positive and decreasing function, this means that, for $s \to \infty$, the *dissipation integral* $\Delta(s)$ has to be bounded. Hence, if the operator $\hat{D}(\vartheta)$ does not disappear too fast at infinity

$$|\dot{\widetilde{V}}(\vartheta)>_{\vartheta \to \infty} \to 0 \tag{5.19}$$

and from (5.7)

$$\hat{A}|\widetilde{V}(\vartheta)>_{\vartheta \to \infty} \to 0 \tag{5.20}$$

We took into account here that (5.19) implicates the asymptotic disappearance of the acceleration vector because

$$|\dot{\widetilde{V}}(s)> - |\dot{\widetilde{V}}(0)> = \int_0^s |\ddot{\widetilde{V}}(t)> dt$$

We understand every vector equality as the equality of their components. Hence, for $s \to \infty$, the above integrals have to be bounded.

Multiplying (5.5) by the operator $\hat{A}$, we get

$$\hat{A}|V(s)> = \hat{A}|\widetilde{V}(s)> + |Z>$$

(5.21)

see (5.4). In other words, the

$$\lim_{s\to\infty}\hat{A}|V(s)> = |Z>$$

For continuous operators

$$\lim_{s\to\infty}\hat{A}|V(s)> = \hat{A}|V(\infty)>$$

(5.22)

Hence, we get equation

$$\hat{A}|V(\infty)> = |Z>$$

(5.23)

which is equivalent to Eq. (3.9). In other words, the solutions of the Newton EE (5.2) asymptotically tend to solutions of Eq. (3.9).

## 5.4 IDEAL CONSTRAINTS

For evolutions with constraints, instead of EE (5.2) we have Eq. (6.7). Then, after substitution (5.5) to (6.7), we get

$$|\ddot{\widetilde{V}}(s)> + \hat{D}(s)|\dot{\widetilde{V}}(s)> + \hat{A}|\widetilde{V}(s)> = |R(s)>$$

(5.24)

see (5.7). Now, repeating (5.14-5) we get

$$\frac{d}{ds}\left(<\dot{\widetilde{V}}|\dot{\widetilde{V}}> + <\widetilde{V}|\hat{A}|\widetilde{V}>\right) = -2<\dot{\widetilde{V}}|\hat{D}(s)|\dot{\widetilde{V}}> + <\dot{\widetilde{V}}|R> + <R|\dot{\widetilde{V}}>$$

(5.25)

From (5.5)

$$|\dot{V}(s)> = |\dot{\widetilde{V}}(s)>$$

(5.26)

Hence, for **scleronomic ideal constraints**, for which the velocity vectors $|\dot{V}(s)>$ belong to the same space as the virtual displacements, Sec.6, the last two terms disappear and as in the free case, the energy (5.17) of the system, due to friction, is a decreasing function of the time $s$.

For the **rheonomic constraints**, due to the work done by the reaction forces $|R(s)>$, the system energy may not be a decreasing function of $s$. We shall assume however that as in the case of scleronomic ideal constraints, **the dissipation integral** $\Delta(s)$ **is bounded**. This means that in both cases (5.19) takes place. Another justification for (5.19) is the following: differentiating the constraint Eq. (6.1) with respect to $s$ we get

$$\left(\dot{\hat{L}}(s) + \dot{\hat{N}}(s)\right)|V(s)> + \left(\hat{L}(s) + \hat{N}(s)\right)|\dot{V}(s)> = 0$$

(5.27)

We choose $s$-dependence of the operators $\hat{L}(s) + \hat{N}(s)$ in such a way that

$$\left(\dot{\hat{L}}(s) + \dot{\hat{N}}(s)\right)|V(s)> \to 0, \quad for\ s \to \infty$$

(5.28)

Hence, for appropriate large *s*

$$\left(\hat{L}(s)+\hat{N}(s)\right)|\dot{V}(s)>\cong 0$$

(5.29)

This means that, for an appropriate large *s*, vectors $|\dot{V}(s)>$ are orthogonal to the reaction forces of the constraints and from (5.26), the last two terms of (5.25) disappear. The energy (5.17), after some time, becomes a decreasing function of *s*. This means that, for $s \to \infty$, the velocity vector $|\dot{\tilde{V}}(s)>$ tends to the zero vector. Hence, the acceleration vector also tends to the zero vector.

Taking scalar product of Eq. (5.24) with the vector $<\tilde{V}(s)|$ we get

$$<\tilde{V}(s)|\ddot{\tilde{V}}(s)>+<\tilde{V}(s)|\hat{D}(s)|\dot{\tilde{V}}(s)>+<\tilde{V}(s)|\hat{A}|\tilde{V}(s)>=<\tilde{V}(s)|R(s)>$$

(5.30)

From (5.5), we have

$$|V(\infty)>=|\tilde{V}(\infty)>+\left(\hat{I}+\hat{L}_R^{-1}\hat{N}\right)_G^{-1}\hat{P}_L|V>$$

(5.31)

The first vector satisfies the original RKLE, see (6.1-2) as does the third one, see considerations after (5.6). Hence, because RKLE are linear, homogenous equations, we conclude that the vector $|\tilde{V}(\infty)>$ also satisfies RKLE (1.4). Because of the ideal constraints, the vectors $|\tilde{V}(\infty)>,|R(\infty)>$ are orthogonal. Hence and from (5.30) we get

$$<\tilde{V}(\infty)|\hat{A}|\tilde{V}(\infty)>=0$$

(5.32)

For a positively defined operator, only the trivial vector, $|\tilde{V}(\infty)>\equiv 0$, satisfies Eq. (5.32) and from (5.31) we see that the vector $|V(s)>$ satisfying EE (5.2) asymptotically tends to a solution of RKLE (1.4). This can be proved with relatively small amount of assumptions in the constraintless case, see (5.8-23) and in the case of scleronomic ideal constraints, see (5.24-6). This proof can be repeated for rheonomic ideal constraints if **we assume** that the dissipation integral $\Delta(s)$, see (5.18), is bounded. For such constraints, in the free Fock space *F*, one can derive closed equations for n-pfs, see Sec. 6 and 7, and in fact, according to author's knowledge, these are essentially new equations in the context of the closure problem.

For rheonomic constraints which tend to the scleronomic constraints

$$\left(\hat{L}(s)+\hat{N}(s)\right)\Rightarrow\left(\hat{L}+\hat{N}\right),\quad for\quad s\to s_0$$

(5.33)

we have for $s > s_0$, the scleronomic constraints

$$\left(\hat{L}+\hat{N}\right)|V(s)>=0$$

(5.34)

In this case, the total energy (5.17), for $s > s_0$, is a decreasing function of the time *s*, see (5.25), and we can conclude as above that RKLE described in (3.9) form, admit an evolutionary description presented below, see also (5.24).

# 6. EVOLUTIONS WITH CONSTRAINTS

Upon the EE (5.1) or (5.2) we impose *constraints* described by the *parameterized* RKLE (2.7)

$$\left(\hat{L}(s)+\hat{N}(s)\right)|V(s)>=0 \tag{6.1}$$

with operators

$$\left(\hat{L}(s)+\hat{N}(s)\right) \Rightarrow \left(\hat{L}+\hat{N}\right), \quad for \quad s \to \infty \tag{6.2}$$

These are linear constraints depending on fictitious time *s*. In fact, we can speak here about a *deformation* of the original RKLE (1.4) with *s* – a parameter of deformation. Moreover, we assume temporarily that parameterized operators have the same projection properties as their primary operators to which they tend, see (2.14-16). Below, we give an example of such constraints.

Introducing and constructing below a right inverse to the operator $\hat{N}(s)$:

$$\hat{N}(s)\left(\hat{N}(s)\right)_R^{-1} = \hat{I} \tag{6.3}$$

we equivalently rewrite (6.1) as follows:

$$\left(\hat{I}+(\hat{N}(s))_R^{-1}\hat{L}(s)\right)|V(s)>=\hat{P}_N(s)|V(s)> \tag{6.4}$$

with the projector

$$\hat{P}_N(s) = \hat{I} - (\hat{N}(s))_R^{-1}\hat{N}(s) \tag{6.5}$$

upon the null space of the operator $\hat{N}(s)$. A general solution to (6.1) or (6.4) can be presented as

$$|V(s)>=\left(\hat{I}+(\hat{N}(s))_R^{-1}\hat{L}(s)\right)^{-1}\hat{P}_N(s)|V(s)> \tag{6.6}$$

where n-pfs generated by the projection vector $\hat{P}_N(s)|V(s)>$ play the role of generalized coordinates. The inverse appearing in (6.6) can be effectively calculated because $(\hat{N}(s))_R^{-1}\hat{L}(s)$ is a product of the lower triangular and diagonal operators, Sec.3.

The rheonomic constraints (6.1) "induce" in, e.g., EE (5.2) the *reaction force* |R>:

$$|\ddot{V}(s)>+\hat{D}(s)|\dot{V}(s)>+\hat{A}|V(s)>-|Z>=|R(s)> \tag{6.7}$$

In the case of ideal constraints, |R> is perpendicular to the *constraint surface* described by (6.6) or (6.1-2). To incorporate the above property of the reaction forces |R> into Eq. (6.7), we introduce the notion of *virtual displacements* $\delta|V>$. Because of the linearity of constraints (6.1-2), these satisfy identical equations

$$\left(\hat{L}(s)+\hat{N}(s)\right)|\delta V(s)>=0 \tag{6.8}$$

(frozen constraint surface (Gantmacher (1966)).

## 6.1 D'ALEMBERT PRINCIPLE

For the *ideal constraints*

$$<\delta V(s)\,|\,R(s)>\, = 0 \tag{6.9}$$

(frozen constraint surface; Gantmacher (1966)). Multiplying (6.7) by the co vector $<\delta V(s)|$ and taking into account (6.9) we get the *d'Alembert principle*

$$<\delta V(s)\,|\,\{|\ddot{V}> + \hat{D}(s)|\dot{V}> + \hat{A}|V> - |Z>\} = 0 \tag{6.10}$$

Goldstein (1975). In the d'Alembert principle dual virtual displacements $<\delta V(s)|$ appear which satisfy the conjugate equation

$$<\delta V(s)\,|\,(\hat{L}*(s) + \hat{N}*(s)) = 0 \tag{6.11}$$

Its general solution can be formally presented as

$$<\delta V(s)\,|\,\hat{P}_N(s)*\left(\hat{I} + \hat{L}(s)*(\hat{N}(s)*)_L^{-1}\right)^{-1} = <\delta V(s)| \tag{6.12}$$

The inverse in this formula is presented in the (formal) Neumann series

$$\left(\hat{I} + \hat{L}(s)*(\hat{N}(s)*)_L^{-1}\right)^{-1} = \hat{I} - \hat{L}(s)*(\hat{N}(s)*)_L^{-1} + \left(\hat{L}(s)*(\hat{N}(s)*)_L^{-1}\right)^2 - \ldots \tag{6.13}$$

in which the left inverse to the operator $\hat{N}(s)*$ appears. This means that this is a series in the inverse powers of the constants $\lambda$ or $\mu$. Unfortunately, equations for n-pfs, obtained from the d'Alembert principle (6.10) with representation (6.12-13) of virtual displacements, still form an infinite chain of equations because of the upper triangular character of the left inverse operator $(\hat{N}(s)*)_L^{-1}$. To obtain closed equations for n-pfs, we use another representation for the virtual displacements, which can be obtained after multiplication of Eq. (6.11) by the operator $(\hat{L}*(s))_L^{-1}$ (inverse operation) and adding to both sides of the obtained equation the same vector with an arbitrary parameter $\alpha$

$$<\delta V(s)\,|\,(\hat{L}*(s)(\hat{L}*(s))_L^{-1} + \hat{N}*(s)(\hat{L}*(s))_L^{-1} + \alpha \cdot \hat{P}_N*(s)) = <\delta V(s)\,|\,\hat{P}_N*(s) \cdot \alpha \tag{6.14}$$

Introducing projector

$$\hat{Q}_L(s) \equiv (\hat{L}(s))_R^{-1}\hat{L}(s) \tag{6.15}$$

and taking into account the formula after (5.4), we describe (6.14) as

$$<\delta V(s)\,|\,(\hat{Q}_L*(s) + \hat{N}*(s)(\hat{L}*(s))_L^{-1} + \alpha \cdot \hat{P}_N*(s)) = <\delta V(s)\,|\,\hat{P}_N*(s) \cdot \alpha \tag{6.16}_a$$

For a right inverse operator $\hat{L}(s)$, these are equivalent equations to Eq. (6.11).

Choosing $\alpha = -2$ and introducing the non-singular *involution operator*

$$\hat{T}_N = \hat{I} - 2\hat{P}_N$$

(reflection at $\hat{P}_N F$ along $(\hat{I} - \hat{P}_N)F$) on vectors decomposed into the direct sum according to the decomposition

$$F = \hat{P}_N F \oplus (I - \hat{P}_N)F$$

see Sec. 4.3) we describe (6.16)$_a$ as

$$<\delta V(s)|\left(\hat{T}_N*(s) - \hat{P}_L*(s) + \hat{N}*(s)(\hat{L}*(s))_L^{-1}\right) = <\delta V(s)|\hat{P}_N*(s) \cdot (-2)$$

(6.16)$_b$

Hence, the dual virtual displacements,

$$<\delta V(s)| = -2 <\delta V(s)|\hat{P}_N*(s)\hat{X}(s)$$

(6.17)$_a$

where the operator

$$\hat{X} = \left(\hat{T}_N*(s) - \hat{P}_L*(s)\right)_R^{-1}\left(\hat{I} + \hat{N}*(s)(\hat{L}*(s))_L^{-1}\left(\hat{T}_N*(s) - \hat{P}_L*(s)\right)_R^{-1}\right)^{-1}$$

(6.17)$_b$

For parameterization of (6.18-19), the difference of involution and projector operator $\left(\hat{T}_N - \hat{P}_L\right)$ is the s-independent, formally Hermitian and diagonal operator ($\hat{N} = \hat{N}_i, i = 1,2.$). The last property means that the inverse to that difference, the existence of which we assume, is also a diagonal operator, Sec.2.

A second inverse operator of (6.17)$_b$ exists and can be effectively constructed as the unit + lower triangular operator.

It turns out that the d'Alembert principle (6.10) together with formulas (6.6) and (6.17) lead to closed equations for n-pfs. As we will see, a closure of RKLE (1.4) depends on those properties of the $\hat{L}(s)$ and $\hat{N}(s)$ operators, see (6.1-2), which allow us to invert the diagonal operator $\left(\hat{T}_N - \hat{P}_L\right)$ in particular n-point subspaces, $n=1,2,3,...,$. As we shall see, in the lowest sub sectors of that operator, where

$$\hat{P}_n\left(\hat{T}_N - \hat{P}_L\right) = -\hat{P}_n\left(\hat{I} + \hat{P}_L\right)$$

(6.18)

for $n=1,2,$ we can construct the inverse, because

$$\left(\hat{I} + \hat{P}_L\right)^{-1} = I - \frac{1}{2}\hat{P}_L$$

(6.19)

## 6.2 PARTICULAR PARAMETRIZATION
We will demonstrate these closed equations in the case of the lowest n-pfs when the rheonomic constraints (6.1) are chosen in such a way that the projectors appearing in the virtual displacements (6.17) do not depend on the "time" $s$. This takes place, for example, for

$$\hat{L}(s) = \lambda(s)\hat{L}, \quad \hat{N}(s) = \nu(s)\hat{N}$$

(6.20)

when

$$\hat{P}_L(s) = \hat{P}_L = \hat{I} - \hat{Q}_L, \quad \hat{P}_N(s) = \hat{P}_N = \hat{I} - \hat{Q}_N$$

(6.21)

Assumptions (6.20) leading to the *s*-independent projectors (6.21) may not be acceptable in the case of an initial theory |V(0)> in which subsequent approximations are not uniformly convergent. In such a case we can use other deformations of the initial theory (1.3-4), see Sec. (8.5) and Nayfeh (1973). This may lead to s-dependence of the projectors $\hat{P}_{L,N}$ and to a more complicated theory.

The most important of the concepts just introduced is that of the *closing* the infinite chain of Eq. (1.4). What we have achieved is to show that, given a certain collection of assumptions, *it is possible to solve the closure problem of RKLE* without the usual explicit or implicit truncation of the infinite chain of equations like (1.4). By an *implicit truncation* we understand a situation in which, for example, a closed equation for a given n-pf is derived with the kernel or other quantity requiring new branching equations to be solved.

We would like to remark that the different evolutionary approximations to a variety of equations from simple algebraic equations to equations like (1.4) are well known in the literature. Linear computers simply like evolutionary formulations, see Na (1979). In the case of quantum field theory such an approach based on the extended Fock space was adopted in a series of author papers, Hańćkowiak (1992a), (1992b), (1993), (1994) where equations for Green's functions were described in the form of RKLE. In the present paper we cast RKLE (1.4) in such an evolutionary form that the resulting equations for n-pfs are **closed,** Sec.7. Moreover, using the free Fock space with Cuntz operators $\hat{\eta}(x), \hat{\eta}^*(y)$, Kozyrev (2004), allows us to construct appropriate right inverse operators and projectors upon the null spaces of the corresponding operators in an explicit way. .

# 7. CLOSED EQUATIONS FOR THE LOWEST N-POINT FUNCTIONS

We consider the so-called $\varphi^4$-theory in which $\lambda = 0$ in (1.1) and RKLE (1.4) or (2.7). The d'Alembert principle (6.10) with the dual virtual displacements (6.17) leads, in fact, to the Lagrange's equations of the second kind, for the *generalized coordinates* represented here by the coefficients of the generalized vector $\hat{P}_N | V >$, considered also in Hańćkowiak (1994) for the Aristotelian evolutions. We consider Eq. (6.10), for the case of $\lambda = 0$, in which, in all derived formulas, the operator $\hat{N} \to \hat{N}_2$, see (2.6). In this case, the projector

$$\hat{Q}_{N2} \equiv \hat{R}_2 \hat{N}_2 = \int N_{2R}^{-1}(y_{(4)}) \hat{\eta}^*(y_1) \cdots \hat{\eta}^*(y_3) \eta(y_4)^3 \cdot dy_{(4)}$$

(7.1)

with function $N_{2R}^{-1}$ satisfying (3.7). This is a Hermitian projector for

$$N_{2R}^{-1} = \delta(y_1 - y_2)\delta(y_2 - y_3)\delta(y_3 - y_4)$$

(7.2)

but in this case (3.7) will be satisfied only for discrete values of space-time variables in which case the Dirac delta are substituted by the Kronecker delta. Then, the projector (5.19)

$$\hat{P}_{N2} = \hat{I} - \hat{Q}_{N2} = \hat{P}*_{N2}$$

(7.3)

is also a Hermitian projector. We therefore assume discretization of the space-time. This means that all integration symbols should be substituted by summations, but here we retain the previously used notation for the sake of simplicity. In this way we can see a significant increase in the set of local operators like

$$\int \hat{\eta}*(x)^2 r(x,y)\hat{\eta}(y)^2 \, dxdy$$

(7.4)

allowing for self multiplications. This and contemporary digitalization techniques suggest that discretization is an appropriate language for describing some aspects of the Nature.

Discretization is also used in Quantum Field Theory to avoid squares of Dirac's delta functions, Weinberg (1996, vol.1).

So, the d'Alembert principle (6.10) with the dual virtual displacements (6.17) and the s-independent Hermitian projectors leads to equations

$$\hat{P}_{N2}\hat{X}(s)\{|\ddot{V}(s)> + \hat{D}(s)|\dot{V}(s)> + \hat{A}|V(s)> - |Z>\} = 0$$

(7.5)

in which the vector |V(s)> satisfying the constraints (6.1-2) is given by (6.6) with the operator $\hat{N} = \hat{N}_2$. In fact, these are equations upon the vector $\hat{P}_{N2}|V(s)>$. Because of the presence of the operator $\hat{X}(s)$, see (6.17)$_b$, multiplied by the projector $\hat{P}_{N2}$ on the null space of the operator $\hat{N}_2$, the above equations differ from Eq. (5.2) considered without any constraints. In some sense, they are similar to the linear differential-algebraic equations

$$Ex'(t) + Fx(t) = q(t)$$

considered by Schulz (2003), where $E, F \in L(R^m)$, $q(t) \in R^n$. In the case of (7.5) in order to get a canonical form of the equation in which the second derivative $\hat{P}_{N2}|\ddot{V}(s)>$ is expressed by the lower derivatives and the lower projections, we need to assume that the operator

$$\hat{P}_N\left(\hat{I} - \hat{P}_L*(s)\hat{T}_N*(s)\right)_R^{-1}\hat{P}_N$$

is invertible or left invertible only on the space $\hat{P}_N F$.

In the lowest section of the free Fock space $F$t for the $\varphi^4$-model

$$\hat{P}_n\hat{P}_{N2} = \hat{P}_n$$

(7.6)

for $n=1,2$. In this case, the first inverse operator appearing in the operator $\hat{X}$, see (6.17)$_b$, projected with these projectors can be inverted and

$$\hat{P}_n\left(\hat{I} + \hat{N}*(s)\left(Q_L*(s) + \alpha \cdot \hat{P}_N*(s)\right)^{-1}\right)^{-1} = \hat{P}_n$$

(7.7)

for $n=1,2$. Hence, we eventually obtain from (7.5)

$$\hat{P}_n \{|\ddot{V}> + \hat{D}(s)|\dot{V}> + \hat{A}|V> - |Z>\} = 0$$

(7.8)

for *n=1,2*. In other words, for low projections, the d'Alembert principle leads to the same equations as the original EE (5.2) - in spite of using constraints (6.1), (6.9)! Now we examine these equations together with formula (6.6), for which

$$\hat{P}_n |V> = \hat{P}_n \hat{P}_N |V>$$

(7.9)

for *n=1,2*. Taking into account that from (5.4)

$$\hat{P}_n \hat{A} = \hat{P}_n (\hat{I} + \hat{L}_R^{-1} \hat{N})$$

(7.10)

for *n=1,2,* we finally get closed equations for the lowest n-pfs:

$$\hat{P}_n \left\{ |\ddot{V}(s)> + \gamma |\dot{V}(s)> + \left( \hat{I} - \frac{\lambda(s)}{\nu(s)} \hat{Q}_L \right) |V(s)> - \hat{P}_L |V> \right\} = 0$$

(7.11)

in which (5.11) and (6.20) were used. In these equations, the projections $\hat{P}_n \hat{P}_L |V>$, for *n=1,2,* are considered as known quantities. In fact they do not depend on the variable *s* and can be calculated from the demand that the n-pfs, for $s \to \infty$, are permutationally symmetric.

We have to solve these equations at the initial condition

$$|\dot{V}(s)>|_{s=0} = |\dot{V}(0)>, \quad |V(s)>|_{s=0} = |V(0)>$$

(7.12)

These two initial vectors have to be chosen in accordance with constraints (6.1).

Eq. (7.11) can be decomposed into the two equations

$$\hat{P}_n \left\{ \hat{Q}_L |\ddot{V}(s)> + \gamma \hat{Q}_L |\dot{V}(s)> + \left( 1 - \frac{\lambda(s)}{\nu(s)} \right) \hat{Q}_L |V(s)> \right\} = 0$$

$$\hat{P}_n \{\hat{P}_L |\ddot{V}(s)> + \gamma \hat{P}_L |\dot{V}(s)> + \hat{P}_L |V(s)> - \hat{P}_L |V>\} = 0$$

(7.13)

where *n=1,2,* and projectors $\hat{Q}_L, \hat{P}_L$ do not depend on the parameter *s*. So we see that the first equation reminds us of the damped harmonic oscillator with the *s*-dependent rigidity coefficient diminishing with growing *s*. The second equation reminds us of a damped harmonic oscillator with constant "force" $\hat{P}_L |V>$ and solutions

$$\hat{P}_L |V(s)>_{s\to\infty} \Rightarrow \hat{P}_L |V>$$

(7.14)

To examine Eq. (7.13)$_1$, we describe this equation as

$$y'' + \gamma \cdot y' + f(s)y = 0$$

(7.15)

with function

$$f(s)|_{s\to\infty} \to 0$$

(7.16)

Choosing function *f(s)* such that a general solution to (7.15) can be explicitly expressed by special or elementary functions we get compact formulas. So, we consider equation (7.15) with function

$$f(s) = b\exp(-2\gamma s)$$

(7.17)

The general solution to this equation is given by a formula

$$y(\xi) = A\exp\left(i\frac{\sqrt{b}}{\gamma}\xi\right) + B\exp\left(-i\frac{\sqrt{b}}{\gamma}\xi\right); \quad \xi = \exp(-\gamma s); \quad \gamma, b > 0$$

(7.18)

Polyanin et al. 2000. The constants *A,B* can be found by means of the initial conditions

$$y(0) = A\exp\left(i\frac{\sqrt{b}}{\gamma}\right) + B\exp\left(-i\frac{\sqrt{b}}{\gamma}\right)$$

(7.19)

$$y'(0) = -i\sqrt{b}\left[A\exp\left(i\frac{\sqrt{b}}{\gamma}\right) - B\exp\left(-i\frac{\sqrt{b}}{\gamma}\right)\right]$$

The final quantity, $y(\xi)|_{s\to\infty}$, is given by

$$y(\infty) = A + B$$

(7.20)

In the general case

$$\vec{y}_0 = T^{-1} \cdot \vec{x}_0$$

(7.21)

where the vectors $\vec{y}_0 = (y(0), y'(0))$, $\vec{x}_0 = (A, B)$ and the inverse matrix $T^{-1}$, in case of (7.19) is

$$T^{-1} = \begin{pmatrix} \exp(i\sqrt{b}/\gamma), & \exp(-i\sqrt{b}/\gamma) \\ -i\sqrt{b}\exp(i\sqrt{b}/\gamma), & i\sqrt{b}\exp(-i\sqrt{b}/\gamma) \end{pmatrix}$$

(7.22)

In the general case

$$\vec{x}_0 = T \cdot \vec{y}_0$$

(7.23)

We call the matrix *T* a *transit matrix*. From (7.20) we get

$$y(\infty) = T_{11}y(0) + T_{12}y'(0) + T_{21}y(0) + T_{22}y'(0)$$

(7.24)

Using the previous notation we get

$$\hat{P}_n|V(\infty)> = \hat{P}_n(\hat{Q}_L + \hat{P}_L)|V(\infty)> \equiv \hat{P}_n|V> =$$
$$\hat{P}_n\{\hat{Q}_L[(T_{11} + T_{21})|V(0)> + (T_{12} + T_{22})|\dot{V}(0)>] + \hat{P}_L|V>\}$$

(7.25)

for *n=1,2*. We bear in mind that the role of projection $\hat{P}_L |V>$ is to make symmetric the expression for a given n-pf. This can be done if this projection

$$\hat{P}_n \hat{P}_L |V> = \hat{P}_n \{\hat{P}_L [(T_{11} + T_{21})|V(0)> + (T_{12} + T_{22})|\dot{V}(0)>]\}$$
(7.26)

and the initial vectors are given by formulas (1.3) with the s- dependent general solutions. Then, from (7.25) and (7.26), we get

$$\hat{P}_n |V> = \hat{P}_n |V(\infty)> = \hat{P}_n \{[(T_{11} + T_{21})|V(0)> + (T_{12} + T_{22})|\dot{V}(0)>]\}$$
(7.27)

for *n=1,2*. It is worth of stressing that (7.27) takes place for any kind of Eq. (7.13), but different coefficients $T_{ij}$ will appear with a different choice of functions *f(s)* or $\lambda(s), \nu(s)$, where

$$f(s) = 1 - \frac{\lambda(s)}{\nu(s)}$$
(7.28)

These functions have to be chosen in such way that the initial vectors $|V(0)>, |\dot{V}(0)>$ can be easily constructed. We can choose, for example, $\lambda(0)/\nu(0) >> 1$, to be able to use the perturbation theory for the construction of the initial vectors. In this case, however, the choice (7.17) leads to negative values of the parameter *b* what is forbidden by assumptions (7.28). In other words, for parameterization (7.17) and (7.18), we should rather take $\lambda(0)/\nu(0) << 1$. This case corresponds to the strong non-linear term in the initial theory, by means of which a theory with a different non-linear term can be expressed by the formula (7.27).

In the case in which matrix elements $T_{ij}$ depend on parameters which should not enter the n-pfs, *n=1,2*, the initial vectors should depend on these parameters in such way that the l.h.s. of (7.27) does not depend on these parameters. In the case of one parameter, $\vartheta$,, by differentiating (7.27) we get

$$\hat{P}_n \left\{ \frac{\partial}{\partial \vartheta} [(T_{11}(\vartheta) + T_{21}(\vartheta))|V(0)> + (T_{12}(\vartheta) + T_{22}(\vartheta))|\dot{V}(0;\vartheta)>] \right\} = 0$$
(7.29)

Such an equation can simplify the calculation of e.g., $|\dot{V}(0;\vartheta)>$, by expressing this by $|\dot{V}(0;\vartheta_0)>$. We assume here that |V(0)> does not depend on $\vartheta$.

# 8. FINAL REMARKS

### 8.1 N-POINT LINEARIZATION

If we believe that the linear property is a recommended feature of any theory, we should appreciate the fact that the statistical description (1.3-4) of the original non-linear equation (1.1) is linear. In some sense, we can regard Eq. (1.4) as a linearization of Eq. (1.1). This linearization is made by substitution of the original 1-pf $\varphi(x)$ by the infinite set of n-pfs $V(x_1,...,x_n) \equiv V(x_{(n)})$, for $n=1,2,...$ which are averaged products of the field $\varphi(x)$. In many cases these n-pfs are more directly related to the experimental than the original 1-pf $\varphi(x)$. Moreover, for "sharp" probability distributions $P$ in Eq. (1.3), one can in principle retrieve $\varphi(x)$ and its products from a knowledge of n-pfs $V(x_{(n)})$.

## 8.2 GENERATING VECTORS AND ANTI-CUNTZ ALGEBRA
In the paper we try to handle an infinite set of equations for n-pfs by introducing the generating vectors (1.7) belonging to the free Fock space $F$ created with the help of the cyclic vector (1.10) and the operators satisfying the anti-Cuntz relations (1.9). In space $F$, the operators related to the linear and non-linear parts of the original theory (1.1) are expressed by the operators satisfying these relations. This facilitates different transformations of equations for n-pfs.

## 8.3. N-POINT SMEARING
Instead of the averaging (1.3) with respect to additional conditions imposed upon solutions of Eq. (1.1), we can use averaging with respect to the $x$-variables. As we know $\varphi(x)$ as well as $\varphi(x-w)$, where $w$ is a vector parameter, satisfy Eq. (1.1) with constant coefficients. These solutions satisfy of course different IBC; for $w = (\alpha, \vec{0})$, the $\alpha$ can related to the initial phase of vibrations. It turns out that moments

$$V(x_{(n)}) = i^n \int \varphi(x_1 - w)...\varphi(x_n - w) W(w) \, dw$$

(8.1)

satisfy exactly the same equations (1.4) as the moments (1.3). The moments (8.1) express the fact that we are not able to prescribe the field to every points of a given region but rather to their neighborhoods described by the weight function $W(w)$. For comparison, see window functions used by Dreyer et al. (1999).

In the case of linear equations, ($\lambda = \mu = 0$), the averaged field satisfies identical equation (1.1). In this case we automatically take into account the averaging procedure by using smeared initial and boundary conditions.

One can consider more general averages then (8.1):

$$V(x_{(n)}) = i^n \int \varphi(x_1 - w_1)...\varphi(x_n - w_n) W_n(w_{(n)}) \, dw \equiv V[x_{(n)}; W_n]$$

(8.2)

In this case, to derive equations for n-pfs like (8.2), we have to consider functions depending on different weight functions $W_n = W(w_{(n)})$. Acting with operator $L$ on the first variable $x_1$ and using (1.1) we get

$$L_1 V[x_{(n)}; W_n] - i\lambda \cdot V[x_1, x_1, x_2,..., x_n; W_{n+1}^2] - \mu \cdot V[x_1, x_1, x_1, x_2,..., x_n; W_{n+3}^3] = 0$$

(8.3)

where the weight functions

$$W_{n+1}^2 = \delta(w_1 - w_2) W_n(w_2,...,w_{n+1}), \quad W_{n+2}^3 = \delta(w_1 - w_2)\delta(w_2 - w_3) W_n(w_3,...,w_{n+2})$$

(8.4)

One can get n-pfs (8.1) and RKLE (1.4) if we use weight functions

$$W_{n+1}^2 = \delta(w_1 - w_2)\cdots\delta(w_n - w_{n+1})W_1(w_{n+1}),\ W_{n+2}^3 = \delta(w_1 - w_2)\cdots\delta(w_{n+1} - w_{n+2})W_1(w_{n+2})$$
(8.5)

For non-symmetrical weight functions $W_n$ the corresponding correlation functions (8.2) are non-symmetrical. For them the free Fock space must be used. Such functions may be related to the different measurement precision of the apparatus located at different space-time.

A simple and perhaps practical example of weight functions leading to non-symmetrical moments (8.2) is the following:

$$W_n(w_{(n)}) = W_1(w_1)\cdots W_n(w_n)$$
(8.6)

### 8.4 DISSIPATION AND RANDOMNIZATION

Recently, Musser (2004), has reminded us of the old idea that beneath Quantum Mechanics there may be a deterministic theory with information loss, see also 't Hooft (2002). It is believed that, due to dissipation, a classical system "may be transformed" into a quantum one which contains less information than the original and which admits a new interpretation of the quantities considered.

Let us look at EE (5.2) or (6.7) in a similar way. They describe linear vibrations with variables $V(s; x_{(n)})$, where $s$ is a time and $x_{(n)} \equiv (x_1,...,x_n)$ enumerates variables $V$. Due to dissipation, with $s \to \infty$, $V(s; x_{(n)})$ tend to $V(\infty; x_{(n)})$ which describes the system in the stationary state. However, according to definition (1.3) and (8.2) they admit a new interpretation: they can be related to the non-linear system (1.3) with random IBC.

A similar idea of obtaining a quantum system (quantum field theory) was described by the author in Hańćkowiak (1992). In this case the complex weight functional in (1.3) has to be used satisfying

$$|P|=1$$
(8.7)

Is this reminiscent of the Laplace's principle of equal ignorance telling us that the probability of each of the outcomes is the same?

### 8.5 OTHER DEFORMATIONS OF EQUATION (1.1)

We assume that constraints (6.1-2) are related to equations

$$L\varphi_s(x) = -\lambda(s)\cdot\varphi_s(f_1(x;s))\cdot\varphi_s(g_1(x;s)) - \mu(s)\cdot\varphi_s(f_2(x;s))\cdot\varphi_s(g_2(x,s))\cdot\varphi_s(h_2(x;s))$$
$$\equiv N(\varphi_s(x);s)$$
(8.7)

where, for $s \to \infty$, the scalar functions $\lambda(s), \mu(s) \to 1$ and the vector functions $f_{1,2}(\ ;s),...,h_2(\ ;s)$ transforming $R^4 \to R^4$ tend to the identity transformations. In other words, Eq. (8.7) substantially generalizing the original Eq. (1.1) tends to Eq. (1.1). Choosing the functions $f_{1,2}(x;s)$ such that in the vicinity of $s = 0$ they take large values, for all $x$, one can see that solutions of the linear equation

$$L\varphi_0(x) = 0$$

(8.8)

which tend to zero, for $|x| \to \infty$, satisfy (8.7) $s \cong 0$. Here, the non-linearity of the r.h.s. of Eq. (8.7) is perfectly utilized. One can hope that for the above deformation, in spite of divergent problems, the initial vector $|V(s)>_{s=0}$ can be effectively described by the free theory (8.8). The vector $|V(\infty)>$, or n-pfs generated by this vector, is related to the vector $|V(0)>$ via the closed equations (7.5).

**8.6 A MORE HOLISTIC APPROACH**

In any theory described by a linear equation with the operator naturally decomposed into two parts, see (2.7), there is a natural tendency to distinguish one of these parts and to solve the simplified theory first. A small constant in front of one of the operators has encouraged such a philosophy. We call such an approach – a perturbation bias.

In the author's opinion, the free Fock space is a possible step to change the above approach; there is no difference in terms of difficulty in inverting the operator $\hat{L}$ related to the linear part and the operator $\hat{N}$ related to the non-linear part of theory (closure problem). We try to solve the problem of too many solutions, which we encounter in the free Fock space, by means of a reformulation of primary equations in an "evolutionary" form with "dissipation". Constraints imposed upon such fictitious evolutions of the system lead to equations upon correlation functions whose closure is due to the appropriate properties of the linear(!) and non-linear parts of the theory. It is remarkable that in order to close equations we only used an assumption about the existence of the inverse operator $(\hat{I} - \hat{P}_L *(s)\hat{T}_N *(s))^{-1}$, see (7.5), in the vicinity of (6.17), and (6.6). Moreover, to get the closed equations for the lowest correlation functions $(\varphi^4)$, we do not need to construct the above inverse.

In particular, via such (closed) equations, theories with various values of the constant(s) can be related. We get here very similar result as in quantum field theory with hidden symmetry, Wal et al. (1997). In our case, the hidden symmetry is the permutation symmetry. It is seen surprisingly that from the point of view of the lowest n-pfs, theories with various values of the constants $\lambda, \mu$ do not differ too much. These results can be viewed as some justification for the philosophical ideas presented by Anselmi (1998) which claim that renormalization and even perturbation have a deep conceptual meanings.

One can also see many common features with the stochastic limit method, Accardi et al. (1999).

Our considerations are quite general and in fact can be applied to many partial differential equations with polynomial non-linearity. The approach presented to the closure problem differs from approaches in which the non-linear terms are substituted by some products of the field averages, Layton et al. (2004).

In the paper, we use the discrete model of space-time, Sec.3.1 as used in the lattice approximations or in the finite element methods We hope that the closed equations derived help us to understand the scale dependence of such theories.